%% file: main.tex
\documentclass[11pt,a4paper]{article}
\usepackage[T1]{fontenc}
\usepackage[utf8]{inputenc}
\usepackage{lmodern,microtype}
\usepackage[margin=25mm]{geometry}
\usepackage{amsmath,amssymb,amsthm,mathtools}
\usepackage{graphicx,placeins,needspace}
\usepackage[hidelinks]{hyperref}
\usepackage{bookmark}
\hypersetup{
  pdftitle={Vanishing Relative Distance of Uniformly Random CSS LDPC Codes Built from Permutation Matrices},
  pdfauthor={Kenta Kasai},
  pdfsubject={Minimum distance and weight distribution of uniformly conditioned CSS codes built from permutation matrices}
}
\allowdisplaybreaks[2]
\numberwithin{equation}{section}
\theoremstyle{definition}

\newtheorem{theorem}{Theorem}[section]
\newtheorem{lemma}[theorem]{Lemma}

\title{Vanishing Relative Distance of\\
Uniformly Random CSS LDPC Codes\\
Built from Permutation Matrices}
\author{Kenta Kasai\\[0.4em]
{\small Institute of Science Tokyo}\\
{\small \href{mailto:kenta@ict.eng.isct.ac.jp}{\texttt{kenta@ict.eng.isct.ac.jp}}}}
\date{}

\begin{document}
\maketitle
\begin{abstract}
Classical random regular low-density parity-check codes can have minimum distance growing linearly with block length.
We study CSS check matrices formed as full $J\times L$ arrays of arbitrary $P\times P$ permutation matrices, sampled uniformly subject to commutation.
For fixed column weight $J\ge3$ and fixed even row weight $L>2J$, we prove that the relative minimum distance converges to zero in probability as $P\to\infty$, with block length $n=LP$.
We analyze the weight enumerators that count all nontrivial logical representatives after excluding stabilizers.
A coefficientwise lower bound shows that, at every fixed relative weight $0<\alpha<1$ along even-weight sequences, exponentially many such representatives exist with high probability.
Their typical and mean counts have exponent lower bound $h_2(\alpha)/2$, where $h_2$ is the binary entropy function.
We compare this bound with the classical LDPC and dense CSS mean weight-distribution exponents.
The existence, with high probability, of a logical operator below every fixed target weight $\delta n$ yields the distance conclusion.
\end{abstract}

\input{introduction.tex}
\input{model_and_distance.tex}
\input{counting_lemmas.tex}
\input{main_theorem.tex}
\appendix
\renewcommand{\thetheorem}{\Alph{section}.\arabic{theorem}}
\renewcommand{\theequation}{\Alph{section}.\arabic{equation}}
\input{counting_appendix.tex}
\input{weight_two_appendix.tex}

\input{references.tex}
\end{document}

%% file: introduction.tex
\section{Introduction: the distance of a uniformly sampled CSS code}
Classical random regular low-density parity-check (LDPC) codes have minimum distance growing linearly with block length, with high probability, for fixed column weight $J\ge3$ and fixed row weight $L>J$~\cite{Gallager1962,Yang2011}.
For Calderbank--Shor--Steane (CSS) codes, the two binary check matrices must also satisfy a commutation condition~\cite{CalderbankShor1996,Steane1996}.
In the dense setting, without a sparsity constraint on the check matrices, random-coding arguments establish the existence of CSS code families with positive rate and linearly growing minimum distance~\cite[Sec.~V]{CalderbankShor1996}.
Structured CSS code families also achieve bounded row and column weights, positive rate, and linearly growing minimum distance~\cite{PanteleevKalachevGood2022,LeverrierZemor2022}.
A natural model closer to permutation-based constructions uses two full $J\times L$ arrays of arbitrary $P\times P$ permutation matrices, with fixed $J,L$ and growing lift size $P$.
Does linear distance growth hold when these lifts are sampled uniformly subject to the CSS commutation condition?

Theorem~\ref{thm:css-uniform} answers this question in the negative.
Fix a column weight $J\ge3$ and an even row weight $L>2J$.
As the lift size $P$ tends to infinity, with block length $n=LP$, the relative minimum distance converges to zero in probability.
The resulting codes all have rate at least $1-2J/L>0$, so the number of logical qubits remains proportional to the block length even as the relative distance vanishes.
The uniform law is obtained by choosing all permutation blocks independently and uniformly, then conditioning on commutation.
It counts labeled matrix pairs, so distinct matrix presentations of the same code are separate samples.
The result distinguishes the existence of structured good CSS code families from the typical behavior of a uniformly sampled matrix pair.

Average weight distributions have been used to study minimum distance and bounds on maximum-likelihood decoding error probabilities since the introduction of classical LDPC ensembles in 1962 and the subsequent 1963 monograph~\cite{Gallager1962}\cite[Sec.~2.2]{Gallager1963}.
The basic object is the expected number of codewords at each weight; its exponential growth rate describes the mean weight spectrum at weights proportional to the block length.
Subsequent asymptotic distance-distribution calculations distinguish the sampling rules of several regular ensembles~\cite{LitsynShevelev2002}, while extensions over finite fields give weight-distribution and minimum-distance results beyond the binary setting~\cite{Yang2011}.
This line of work provides the classical comparison in Section~\ref{sec:classical-spectrum}.

Sparse-graph quantum codes, including bicycle constructions, established early ways to combine sparse checks with commutation~\cite{MacKayMitchisonMcFadden2004}.
Early investigations of the symmetric setting $H_X=H_Z$ also considered random generation and approximate counting of sparse self-orthogonal check matrices, low-weight logical operators, and competing conjectures about bounded and linearly growing quantum distance~\cite[Secs.~4.2--4.4 and~7.4]{MacKayMitchisonMcFaddenExtended2004}.
These observations and conjectures motivate asking about typical distance under a specified distribution on commuting matrix pairs.
Regular quasi-cyclic constructions made the row and column weights explicit design parameters~\cite{HagiwaraImai2007}, while hypergraph-product codes provided positive rate and distance proportional to the square root of the block length~\cite{TillichZemor2014}.
More recent bivariate bicycle codes connect finite-length algebraic design to fault-tolerant quantum memory~\cite{Bravyi2024}.
These developments motivate studying both the structure that enforces commutation and the distance of the resulting code.

The author's affine-permutation and pair-partition (PP) design programs were motivated by the expectation that random CSS codes built from permutation matrices would have linearly growing distance, by analogy with classical random LDPC codes.
The aim was to retain the flexibility of arbitrary permutation matrices while imposing tractable constraints that enforce commutation.
This was a design motivation, rather than an established distance theorem for the conditioned random ensemble.
The present result tests that expectation for the uniform law defined in Section~\ref{sec:uniform-ensemble}.
A recent sequence of regular CSS constructions pursues this problem through controlled commutation of affine permutation matrices~\cite{KasaiOrthogonality2026}, affine-coset bases~\cite{OkadaKasaiAffine2026}, square-base hypergraph products with circulant lifts~\cite{OkadaKasaiSquare2026}, and two-branch finite-field bases~\cite{OkadaKasaiTwoBranch2026}.
The PP construction imposes linear equations on circulant shifts so that terms in each check product cancel in pairs~\cite[Secs.~II--IV]{OkadaKasaiPP2026,LeeEtAlDesign2026}.
The tradeoffs among column weight, check weight, distance, rate, and block length have also been analyzed~\cite{LeeEtAlDesign2026}.

Distance verification is a separate step in this construction program.
Explicit non-stabilizer vectors give certified upper bounds for affine-permutation codes~\cite{KasaiWitness2026}, and complete support searches give lower bounds for fixed PP lifts~\cite[Sec.~V-A]{OkadaKasaiPP2026}\cite{LeeEtAlDesign2026}.
Quaternary coefficient assignments to PP checks also yield complete canonical logical bases after binary expansion~\cite[Secs.~2.3--2.5]{OkadaKasaiLogical2026}.
These results make the distinction between check-kernel vectors, stabilizers, and nontrivial logical representatives explicit, as required by the stabilizer definition of quantum distance~\cite[Secs.~3.2--3.4]{Gottesman1997}.
The same distinction is essential here: counting vectors in a check kernel gives a distance upper bound only after some of them are shown to lie outside the stabilizer space.

Large minimum distances at finite block lengths motivate both affine-permutation codes and PP codes.
For the former, the cited study reports search-based evidence of large distance~\cite{KasaiOrthogonality2026}; for the latter, selected instances also have certified minimum distances~\cite{OkadaKasaiPP2026,LeeEtAlDesign2026}.
These finite-length results do not establish asymptotic linear distance growth.
Even if a suitable sequence of affine-permutation codes or generalized PP codes~\cite[Sec.~III.3]{LeeEtAlDesign2026} were shown to satisfy $d_n\ge c n$ for some constant $c>0$, it would be consistent with Theorem~\ref{thm:css-uniform}.
If its matrix pairs belong to $\mathcal G_P(J,L)$ with the fixed degrees and full base of the theorem, their fraction among the admissible matrix pairs must tend to zero, but they need not cease to exist.
A construction can select such an exceptional family through additional algebraic or graph constraints.
Here generalized PP codes include affine-permutation (APM-PP) variants; the fixed-column-weight CPM-PP family has a separate lift-size-independent distance bound under its stated hypotheses~\cite[Theorem~3]{OkadaKasaiPP2026}\cite[App.~A]{LeeEtAlDesign2026}.

The dependence on the construction is already visible in classical coding.
For a full $J\times L$ array of circulant permutation matrices with $J<L$, the minimum distance is at most $(J+1)!$, independently of the lift size~\cite[Corollary~9]{SmarandacheVontobel2012}.
A recent analysis proves attainment for sufficiently large lifts, both by an explicit construction and with high probability under independent uniform exponent choices~\cite[Secs.~III--IV]{KasaiCPMDistance2026}.
In the quantum setting, a specified family of affine-in-index lifts of a two-branch base has a constant distance upper bound, with an explicit rate-$2/3$ example attaining it~\cite{OkadaKasaiRate2026}.
Thus randomness within a chosen matrix format has to be analyzed for that format.
Related fixed-degree distance results for nested Hsu--Anastasopoulos/MacKay--Neal CSS pairs concern sparse representations with hidden variables; their compressed check matrices are generally dense~\cite[Secs.~2.2 and~2.4]{KasaiFiniteDegree2026}.

Random construction methods address how to generate sparse commuting checks.
Commutation and degree constraints can be imposed by solving a constraint satisfaction problem on random bipartite graphs; the resulting codes have been evaluated numerically on the erasure channel~\cite[Secs.~3--5]{TremblayDuclosCianciKourtis2023}.
A sequential sampler based on information-set decoding offers another way to generate sparse self-orthogonal matrices~\cite{Santini2026}.
In the extended analysis, output uniformity remains unproved, column weights are not fixed, and the independent-row weight-enumerator model is applied to the correlated self-orthogonal matrices as a heuristic~\cite[extended version, Secs.~I.B and~IV.B, Heuristic~3]{Santini2026}.
To preserve both row and column weights, random swaps can be followed by integer-programming repair of the commutation condition; this procedure does not establish a uniform output law on all admissible pairs~\cite[Sec.~II]{OkadaKasaiRandom2025}.

A different random ensemble has a proved positive relative distance.
Combining an asymptotically good quantum LDPC base code, expander-based distance amplification, and independent random inner encodings yields constant-check-weight stabilizer codes approaching the Gilbert--Varshamov bound with high probability~\cite[Theorem~1.2 and Sec.~1.1]{MittalEtAl2026}.
The structured sampling law in that construction shows why the distribution, as well as sparsity and commutation, matters for a distance theorem.

Weight-distribution analysis connects these ensemble questions to distance.
The classical regular weight-distribution exponent has been used as a heuristic distance scale for finite quantum constructions~\cite[Sec.~V.2 and App.~E.3]{LeeEtAlDesign2026}.
The present contribution concerns the conditional uniform distribution on pairs of permutation lifts of the all-ones $J\times L$ base that satisfy the CSS commutation condition, with the fixed degrees specified above.
For this law, we prove a high-probability lower bound on the weight-distribution exponent after excluding stabilizers, and use it to establish vanishing relative minimum distance in Theorem~\ref{thm:css-uniform}.
The conclusion is specific to this uniform ensemble; a construction with a different sampling law can have a different typical distance.

Section~\ref{sec:model} defines the ensemble, CSS distance, and the full weight enumerators of nontrivial logical representatives.
Section~\ref{app:counting} gives a coefficientwise lower bound for these polynomials, derives a lower bound on their weight-distribution exponents, and compares it with the classical LDPC and dense CSS mean exponents.
Section~\ref{sec:main-theorem} uses the high-probability coefficient count to prove the main theorem excluding linear distance growth.
Appendix~\ref{app:enumerator-proof} supplies the combinatorial proof of the coefficient bound.
Appendix~\ref{app:weight-two-proof} retains an alternative proof using weight-two logical operators.

%% file: model_and_distance.tex
\section{The ensemble, CSS distance, and logical weight enumerators}
\label{sec:model}
We define the uniform sampling model, CSS distance, and the weight enumerators of nontrivial logical representatives.
A positive coefficient at a given weight certifies a logical operator of that weight and therefore gives an upper bound on the minimum distance.

\subsection{The uniform ensemble}
\label{sec:uniform-ensemble}
Fix integers $J\ge3$ and even $L\ge2J$.
For a positive integer $P$, set $n=LP$ and $m=JP$.
Let $\mathcal P_P$ be the set of binary $P\times P$ permutation matrices, each having exactly one nonzero entry in every row and column.
There are $P!$ such matrices.
Every binary permutation matrix is allowed as a block, without a circulant or affine restriction.
Define the set of full block arrays and the commuting-pair sample space by
\begin{equation}
 \begin{aligned}
 \mathcal B_P(J,L)&:=
 \{(T_{i\ell})_{1\le i\le J,\,1\le\ell\le L}:T_{i\ell}\in\mathcal P_P\},\\
 \mathcal G_P(J,L)&:=
 \{(H_X,H_Z)\in\mathcal B_P(J,L)^2:H_XH_Z^{\mathsf T}=0\}.
 \end{aligned}
 \label{eq:model-main}
\end{equation}
Each array is interpreted as a binary $m\times n$ matrix.
It is a $P$-fold permutation lift of the all-ones $J\times L$ base matrix, in the protograph sense~\cite[Sec.~II, Definition~1]{Thorpe2003}.
Thus every column has weight $J$ and every row has weight $L$.
The CSS commutation condition couples the two lifts~\cite{CalderbankShor1996,Steane1996}.

Write $U_n$, with $n=LP$, for the uniform distribution on $\mathcal G_P(J,L)$.
Equivalently, choose all $2JL$ blocks independently and uniformly from $\mathcal P_P$, then condition on $H_XH_Z^{\mathsf T}=0$.
This gives the stated uniform law because the fixed block decomposition uniquely determines every block assignment, so all unconditioned matrix pairs have the same probability.
Block indices, row labels, and column labels are fixed; distinct matrix presentations of the same code are separate samples.
Repeated columns and linearly dependent rows are allowed, with no additional girth or column-distinctness condition.
For fixed $H_Z$, its marginal probability is proportional to the number of compatible $H_X\in\mathcal B_P(J,L)$.
The blocks are therefore generally dependent after conditioning.
Lemma~\ref{lem:paired-ensemble-lower} proves nonemptiness for every $P$.

All matrix operations are over $\mathbb F_2$.
We write $\log$ for the natural logarithm and $\log_2$ for the base-two logarithm.
An event holds \emph{with high probability} if its probability tends to one as $P\to\infty$, equivalently as $n=LP\to\infty$, with $J,L$ fixed.
The distance and weight-distribution results assume $L>2J$.
They concern this all-ones base and this conditional uniform law; no claim is made for arbitrary base protographs.

\subsection{Kernel vectors and logical operators}
\label{app:distance-foundations}
Write $\operatorname{row}H$ for the row space and $\ker H=\{v:Hv=0\}$ for the kernel.
The support and weight of a vector are $\operatorname{supp}v=\{i:v_i\ne0\}$ and $\operatorname{wt}(v)=|\operatorname{supp}v|$.
Set $S_X=\operatorname{row}H_X$ and $S_Z=\operatorname{row}H_Z$.
Commutation gives $S_X\subseteq\ker H_Z$ and $S_Z\subseteq\ker H_X$.
The number of logical qubits is $k=n-\operatorname{rank}H_X-\operatorname{rank}H_Z$.
If $L>2J$, then $k\ge n-2m>0$.
We assume $k>0$ whenever discussing distance.

Define the minimum weights of nontrivial $X$-type and $Z$-type logical operators by
\begin{equation}
 d_X=\min_{v\in\ker H_Z\setminus S_X}\operatorname{wt}(v),
 \qquad
 d_Z=\min_{v\in\ker H_X\setminus S_Z}\operatorname{wt}(v).
 \label{eq:logical-distance-definition}
\end{equation}
A vector that satisfies the checks but belongs to the stabilizer space is excluded from the distance minimization.

The minimum Pauli distance of the CSS code is~\cite[Secs.~3.2 and~3.4]{Gottesman1997}
\begin{equation}
 d=\min(d_X,d_Z).
 \label{eq:distance-main}
\end{equation}
Hence the weight of any nontrivial $X$-type or $Z$-type logical operator is an upper bound on $d$.

\subsection{The logical weight enumerators}
\label{sec:logical-enumerators}
For a binary linear space $C\subseteq\mathbb F_2^n$, its weight enumerator is the polynomial
\[
 W_C(z):=\sum_{v\in C}z^{\operatorname{wt}(v)},
\]
where $z$ is a formal variable.
Define
\begin{equation}
 A_{X,n}(w)=
 \#\{v\in\ker H_Z\setminus S_X:\operatorname{wt}(v)=w\}
 \label{eq:logical-enumerator}
\end{equation}
and define $A_{Z,n}(w)$ by interchanging $X$ and $Z$.
These counts distinguish vectors on the physical qubits, including vectors that have the same logical action modulo stabilizers.
Commutation ensures that the stabilizer space is contained in the corresponding kernel, so the polynomials counting only nontrivial representatives are
\begin{equation}
 \begin{aligned}
 W_{X,n}(z)&:=\sum_{w=0}^n A_{X,n}(w)z^w
             =W_{\ker H_Z}(z)-W_{S_X}(z),\\
 W_{Z,n}(z)&:=\sum_{w=0}^n A_{Z,n}(w)z^w
             =W_{\ker H_X}(z)-W_{S_Z}(z).
 \end{aligned}
 \label{eq:logical-polynomials}
\end{equation}
All coefficients are nonnegative, and the constant coefficients are zero.
For two polynomials $F,G$, write $F\succeq G$ if every coefficient of $F$ is at least the corresponding coefficient of $G$.
Thus a coefficientwise lower bound for $W_{X,n}$ or $W_{Z,n}$ directly bounds the number of nontrivial logical representatives at each weight.

To prove the main theorem, it suffices to show that, for every fixed $\delta>0$, a coefficient at some weight below $\delta n$ is positive with high probability.
Section~\ref{app:counting} establishes a stronger statement: at every fixed relative weight $0<\alpha<1$, along even weights, the number of representatives is exponentially large with high probability.
The subtraction in~\eqref{eq:logical-polynomials} is essential: a kernel enumerator alone also counts stabilizers and does not determine a quantum distance upper bound.

%% file: counting_lemmas.tex
\section{Lower bounds on the logical weight distribution}
\label{app:counting}
Fix $J\ge3$, even $L>2J$, and $n\in L\mathbb N$, and sample the check-matrix pair from $U_n$.
The goal of this section is a lower bound on the exponential growth of $A_{X,n}(w_n)$ and $A_{Z,n}(w_n)$ at the even weights $w_n=2\lfloor\alpha n/2\rfloor$, with $0<\alpha<1$ fixed.
We show that $h_2(\alpha)/2$, where $h_2$ is binary entropy, is a lower bound both with high probability and for the exponent of the mean count.

Two lemmas provide the weight-distribution bounds.
Lemma~\ref{lem:enumerator-bound} gives a positive lower bound for every even-weight coefficient from $2$ through a cutoff $w_{\max,n}=2\lfloor\lambda n\rfloor-2$, where $J/L<\lambda<1/2$.
Here $w_n$ is the weight we wish to study, whereas $w_{\max,n}$ is the upper endpoint of the range covered by the lower bound.
The parameter $\lambda$ controls this range: $w_{\max,n}/n\to2\lambda$, so choosing $2\lambda>\alpha$ places $w_n$ within it for all sufficiently large $n$.
Lemma~\ref{lem:logical-spectrum} extracts the coefficient at $w_n$ and evaluates its exponential rate for fixed $\lambda$.
Choosing $\lambda$ sufficiently close to $1/2$ gives the claimed lower bound to any prescribed accuracy.
Section~\ref{sec:classical-spectrum} recalls the classical mean exponent for comparison in Figure~\ref{fig:weight-exponents}.
Section~\ref{sec:main-theorem} then fixes $\alpha$ below any prescribed target relative distance and uses the high-probability lower bound to obtain a nontrivial logical operator.

\subsection{A coefficientwise lower bound}
The first lemma gives a finite-length coefficient bound on an event whose probability tends to one.
Its combinatorial proof in Appendix~\ref{app:enumerator-proof} uses column-multiplicity counting and explicitly excludes stabilizers.

\begin{lemma}[A lower bound on the logical weight enumerators]
\label{lem:enumerator-bound}
Fix $J\ge3$, even $L>2J$, and $J/L<\lambda<1/2$, and put $w_{\max,n}=2\lfloor\lambda n\rfloor-2$ for $n\in L\mathbb N$.
For all sufficiently large $n$, the polynomial on the right below has nonzero coefficients at the even weights from $2$ through $w_{\max,n}$.
With high probability under $U_n$, the inequalities
\begin{equation}
 \begin{aligned}
 W_{X,n}(z)&\succeq z^2(1+z^2)^{w_{\max,n}/2-1},\\
 W_{Z,n}(z)&\succeq z^2(1+z^2)^{w_{\max,n}/2-1}
 \end{aligned}
 \label{eq:enumerator-coefficient-bound}
\end{equation}
hold simultaneously.
\end{lemma}
\begin{proof}
Appendix~\ref{app:coefficient-bound-proof} proves the simultaneous coefficient bound from the commuting-matrix counts and the fixed-weight stabilizer-exclusion argument in Appendix~\ref{app:enumerator-proof}.
\end{proof}

\input{spectrum_comparison.tex}
\FloatBarrier

%% file: spectrum_comparison.tex
\subsection{A lower bound on the weight-distribution exponent}
\label{sec:logical-spectrum}
We now evaluate the exponential growth rate of the coefficients supplied by Lemma~\ref{lem:enumerator-bound}.
The resulting bound concerns the full logical enumerators defined in Section~\ref{sec:logical-enumerators}.

\begin{lemma}[A lower bound on the logical weight-distribution exponent]
\label{lem:logical-spectrum}
Consider the uniform ensemble with fixed $J\ge3$, fixed even $L>2J$, and $n=LP$ as $P\to\infty$.
Fix $0<\alpha<1$ and put $w_n=2\lfloor\alpha n/2\rfloor$.
Let $h_2(\alpha)=-\alpha\log_2\alpha-(1-\alpha)\log_2(1-\alpha)$ be binary entropy.
For every fixed $\varepsilon>0$, with high probability under $U_n$,
\begin{equation}
 \frac1n\log_2 A_{X,n}(w_n)
 \ge\frac12h_2(\alpha)-\varepsilon.
 \label{eq:logical-spectrum-typical}
\end{equation}
Here $\log_2 0=-\infty$.
The mean count also satisfies
\begin{equation}
 \liminf_{n\to\infty}\frac1n\log_2\mathbb E_{U_n}A_{X,n}(w_n)
 \ge\frac12h_2(\alpha).
 \label{eq:full-spectrum-lower}
\end{equation}
The same conclusions hold for the $Z$-type counts.
\end{lemma}
\begin{proof}
To apply the coefficient bound at the target weight $w_n$, choose the cutoff above that weight.
Set $r_n=w_n/2$ and choose a fixed parameter
$\max\{J/L,\alpha/2\}<\lambda<1/2$.
With $w_{\max,n}=2\lfloor\lambda n\rfloor-2$, the inequalities $1\le r_n\le w_{\max,n}/2$ hold for all sufficiently large $n$.
Taking the coefficient of $z^{w_n}$ in Lemma~\ref{lem:enumerator-bound} gives
$A_{X,n}(w_n)\ge\binom{w_{\max,n}/2-1}{r_n-1}$ with high probability.
Stirling's formula~\cite[Prop.~B.1]{FlajoletSedgewick2009} yields
\begin{equation}
 \frac1n\log_2\binom{w_{\max,n}/2-1}{r_n-1}
 \longrightarrow\lambda h_2\!\left(\frac{\alpha}{2\lambda}\right).
 \label{eq:coefficient-entropy-limit}
\end{equation}
As $\lambda$ approaches $1/2$ from below, the right-hand side tends to $h_2(\alpha)/2$.
For a given $\varepsilon>0$, choose $\lambda$ sufficiently close to $1/2$ that this limit exceeds $h_2(\alpha)/2-\varepsilon/2$, and then hold $\lambda$ fixed while $n\to\infty$.
The coefficient lower bound gives~\eqref{eq:logical-spectrum-typical} with high probability.

On the event where~\eqref{eq:logical-spectrum-typical} holds,
$A_{X,n}(w_n)\ge2^{n(h_2(\alpha)/2-\varepsilon)}$.
Since the event has probability tending to one, the mean has lower exponential rate at least $h_2(\alpha)/2-\varepsilon$.
Letting $\varepsilon$ tend to zero gives~\eqref{eq:full-spectrum-lower}.
The coefficient bound for $Z$ gives the same conclusions.
\end{proof}

The lemma gives a lower bound, rather than an exact exponent, for the full logical weight distribution.
It establishes exponentially many nontrivial representatives at every fixed $0<\alpha<1$, along the stated even-weight sequence.
In this model, every vector in either check kernel has even weight for every $J$: summing the rows in any one block row gives the all-ones vector.

\subsection{Comparison with classical LDPC and dense CSS ensembles}
\label{sec:classical-spectrum}
For comparison, use the Gallager regular ensemble~\cite{Gallager1962} with the same degrees $(J,L)$.
Fix an $n/L\times n$ matrix with column weight one and row weight $L$.
Apply independent uniform column permutations to $J$ copies of this matrix and stack the results vertically.
The resulting parity-check matrix has column weight $J$ and row weight $L$.
Denote this random parity-check matrix by $H_{\mathrm{cl}}$ and define
\[
 A_{\mathrm{cl},n}(w)
 :=\#\{v\in\mathbb F_2^n:H_{\mathrm{cl}}v=0,\ \operatorname{wt}(v)=w\}
\]
to be the number of its weight-$w$ codewords.

Fix $J,L$, with $L$ even and $n\in L\mathbb N$.
For an even-weight sequence satisfying $w_n/n\to\alpha\in(0,1)$,
the average weight-enumerator asymptotic for this ensemble~\cite[Sec.~2.2, Eqs.~(2.15)--(2.20)]{Gallager1963} gives, with base-two logarithms,
\begin{align}
 g_{\mathrm{cl}}(\alpha)
 &:=\lim_{n\to\infty}\frac1n\log_2\mathbb E A_{\mathrm{cl},n}(w_n)
 \notag\\
 &=(1-J)h_2(\alpha)
   +\inf_{z>0}\left\{\frac JL\log_2 f_L(z)-J\alpha\log_2 z\right\},
 \label{eq:classical-exponent}\\
 f_L(z)&=\sum_{\substack{0\le k\le L\\k\ \mathrm{even}}}\binom Lk z^k
       =\frac{(1+z)^L+(1-z)^L}{2}.
 \notag
\end{align}
The socket ensemble has the same asymptotic exponent~\cite[accepted manuscript, Thm.~3.6 and Rem.~3.7]{Yang2011}, although its finite-length mean enumerator differs from that of the Gallager ensemble.

A dense CSS ensemble gives a second comparison at the same check-row ratio $m/n=J/L<1/2$.
Choose uniformly an ordered pair of $m$-dimensional subspaces $(S_X,S_Z)$ of $\mathbb F_2^n$ satisfying $S_X\subseteq S_Z^\perp$.
Its rate is exactly $1-2J/L$, the design-rate lower bound of the sparse ensemble; this does not match the actual rate of every sparse sample.
Let $A^{\mathrm{dense}}_{X,n}(w)$ count the weight-$w$ vectors in $S_Z^\perp\setminus S_X$.
The spaces $S_Z^\perp$ and $S_X$ are marginally uniform of dimensions $n-m$ and $m$, respectively.
A fixed nonzero vector belongs to a uniform $t$-dimensional subspace with probability $(2^t-1)/(2^n-1)$, so
\[
 \mathbb E A^{\mathrm{dense}}_{X,n}(w)
 =\binom nw\frac{2^{n-m}-2^m}{2^n-1},\qquad 1\le w\le n.
\]
For $w_n/n\to\alpha\in(0,1)$, this gives
\[
 g_{\mathrm{dense}}(\alpha)
 :=\lim_{n\to\infty}\frac1n\log_2\mathbb E A^{\mathrm{dense}}_{X,n}(w_n)
 =h_2(\alpha)-J/L.
\]
The same formula holds for $Z$; no even-weight restriction is needed.
Let $\delta_{\mathrm{dense}}\in(0,1/2)$ satisfy $h_2(\delta_{\mathrm{dense}})=J/L$.
Writing $d_n^{\mathrm{dense}}$ for the dense code distance, summing the exact mean over both types and $1\le w\le\delta n$ shows that $d_n^{\mathrm{dense}}>\delta n$ with high probability for every fixed $0<\delta<\delta_{\mathrm{dense}}$.
This provides a linear-distance comparison without asserting an exact limiting relative distance.

\begin{figure}[htbp]
 \centering
 \includegraphics[width=\linewidth]{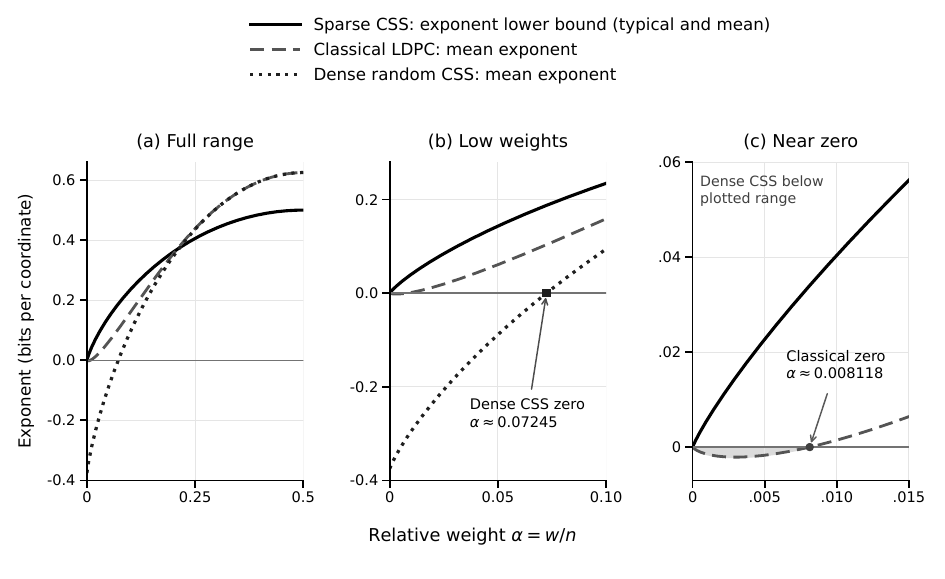}
 \caption{Weight-distribution exponents for $J=3,L=8$, using base-two logarithms.
 The solid curve is the lower bound $h_2(\alpha)/2$ for the sparse CSS logical-representative exponent, valid both typically and in the mean.
 The dashed curve is the classical Gallager mean exponent $g_{\mathrm{cl}}(\alpha)$; the dotted curve is the dense CSS mean exponent $h_2(\alpha)-3/8$.
 At half weight, the two mean exponents are $5/8$, while the sparse lower bound is $1/2$; these heights do not give the quantum rate.
 The $X$- and $Z$-type CSS counts have identical bounds or means.
 Classical and dense first positive zeros are approximately $0.008118$ and $0.072450$, respectively.
 Panel (c) enlarges the region near the classical zero; the dense curve lies below its vertical range.
 The sparse CSS and classical limits use even weights. Values at $\alpha=0$ are continuous extensions, not weight-zero logical counts.
 The classical ensemble matches the sparse degrees; dense CSS matches $m/n=3/8$ and the design rate $1/4$, not necessarily the actual sparse rate.}
 \label{fig:weight-exponents}
\end{figure}

Figure~\ref{fig:weight-exponents} compares the two mean exponents with the sparse CSS lower bound for $(J,L)=(3,8)$.
At relative weight $\alpha=1/2$, the classical and dense CSS mean exponents both equal $1-J/L$.
Indeed, the minimizer in~\eqref{eq:classical-exponent} is $z=1$ at this weight, where $f_L(1)=2^{L-1}$, and $g_{\mathrm{dense}}(1/2)=h_2(1/2)-J/L=1-J/L$.
The common height reflects the check-row ratio $m/n=J/L$.
In the dense CSS ensemble, the total number of nontrivial $X$-type physical representatives is $2^{n-m}-2^m$; since $n>2m$, subtracting the stabilizers does not change its exponential rate $1-m/n$.
These representatives include all $2^m$ members of each nonzero logical coset, so their exponent at half weight is $(1+R_{\mathrm{dense}})/2$, with $R_{\mathrm{dense}}=1-2m/n$, rather than the quantum rate itself.
The classical value $1-J/L$ is its design rate; equality of the plotted heights therefore does not mean equality of classical and quantum code rates.

The sparse CSS curve instead plots the proved lower bound $h_2(\alpha)/2$, whose maximum is $1/2$.
Its factor $1/2$ comes from the approximately $n/2$ disjoint equal-column pairs used to supply logical candidates in Appendix~\ref{app:enumerator-proof}.
The full logical enumerator can also contain representatives outside this counted set.
Thus the lower plotted peak does not establish a smaller maximum of the full sparse CSS exponent or determine the sparse quantum rate.
For $(J,L)=(3,8)$, the classical and dense peaks are $5/8$, the sparse lower-bound peak is $1/2$, and the dense quantum rate is $1/4$.

For the classical ensemble, the mean exponent is negative at small positive relative weights.
At any such fixed relative weight, the expected number of codewords tends to zero; the probability that even one exists is bounded by this expectation.
Together with control over all smaller weights~\cite{Gallager1962,Yang2011}, this shows that the relative minimum distance exceeds any fixed value below the first positive zero with high probability.

For the sparse CSS ensemble, the lower bound $h_2(\alpha)/2$ is positive at every fixed $0<\alpha<1$.
The existence of logical representatives follows from the high-probability bound in Lemma~\ref{lem:logical-spectrum}: their number is exponentially large in a typical sample.
This is the bound used to prove the distance upper bound in Section~\ref{sec:main-theorem}.

To identify a zero of the mean exponent with a limiting relative minimum distance, one also needs a matching high-probability existence result just above that zero.
A positive mean exponent alone can arise from rare samples and does not provide this existence result.

%% file: main_theorem.tex
\Needspace{14\baselineskip}
\section{Main theorem: absence of linear minimum-distance growth}
\label{sec:main-theorem}
We use the preceding count to show that the relative minimum distance of a uniformly sampled code converges to zero in probability.
For any fixed positive target relative distance, a nontrivial logical operator with smaller relative weight exists with high probability, yielding the main theorem.

\begin{theorem}[Absence of linear distance growth in the uniform ensemble]
\label{thm:css-uniform}
Fix an integer $J\ge3$ and an even integer $L>2J$, and put $n=LP$.
Choose $(H_X,H_Z)$ uniformly from the ensemble $\mathcal G_P(J,L)$ in~\eqref{eq:model-main}, and let $d_n$ be the minimum distance of the resulting CSS code.
Then, as $P\to\infty$, for every fixed $\delta>0$,
\begin{equation}
 \Pr_{U_n}\{d_n\ge\delta n\}\longrightarrow0,
 \qquad\text{equivalently,}\qquad
 \frac{d_n}{n}\xrightarrow{\Pr}0.
 \label{eq:uniform-quantum-distance}
\end{equation}
\end{theorem}
\begin{proof}
Fix any target relative distance $\delta>0$, however small.
We will find a logical representative below weight $\delta n$ with high probability.
Set $\alpha=\min\{\delta/2,1/4\}$, so $0<\alpha<\delta$.
The even weight $w_n=2\lfloor\alpha n/2\rfloor$ satisfies
\begin{equation}
 \frac{w_n}{n}\longrightarrow\alpha>0,\qquad w_n<\delta n.
 \label{eq:logical-weight-choice}
\end{equation}
The weight still grows linearly with $n$, but its proportionality constant $\alpha$ has been chosen below the arbitrary target $\delta$.

Apply Lemma~\ref{lem:logical-spectrum} with $\varepsilon=h_2(\alpha)/4>0$.
With high probability, $A_{X,n}(w_n)\ge2^{n h_2(\alpha)/4}>0$, so a nontrivial logical representative of weight $w_n$ exists.
The distance identity~\eqref{eq:distance-main} then gives $d_n\le w_n<\delta n$ with high probability.
The argument applies to every fixed $\delta>0$, proving the conclusion.
\end{proof}

The author initially expected random CSS codes built from permutation matrices to have linearly growing minimum distance, as classical random regular LDPC codes do.
That expectation fails for the ensemble studied here, which samples uniformly from commuting permutation lifts of the all-ones base.
Its relative minimum distance converges to zero in probability: for every fixed positive target relative distance, the probability of attaining it tends to zero.

%% file: counting_appendix.tex
\section{Combinatorial proof of the coefficient bound}
\label{app:enumerator-proof}
This appendix proves Lemma~\ref{lem:enumerator-bound} by counting commuting matrix pairs and excluding stabilizers from fixed-weight sets of kernel vectors.
The underlying structure is concentration on column vectors occurring exactly twice.
We first give the deterministic fixed-weight count, then prove this concentration, and finally obtain the coefficientwise polynomial bound used in the main text.

Throughout the matrix-pair counting argument, fix an integer $J\ge3$ and an even integer $L\ge2J$, and write $n=LP$ and $m=JP$.
All matrix pairs whose probabilities we estimate belong to the sample space $\mathcal G_P(J,L)$ in~\eqref{eq:model-main}.
The counting estimates hold also for $L=2J$; the final coefficient bound uses $L>2J$.
Under the uniform distribution on $\mathcal G_P(J,L)$, the probability of a given $H_Z$ is proportional to the number of compatible matrices $H_X$.
A row of $H_X$ must select $L$ columns of $H_Z$ whose sum is zero, with one selected position in each block column.
Equal columns cancel over the binary field and thus provide simple zero-sum subsets.
Requiring many such pairs also restricts the choices for $H_Z$ itself.
We therefore count both choices together to determine the typical structure.
The denominator is bounded below by an explicit family of permutation lifts.
For exceptional pairs, the one-entry-per-block-row property gives a common parity argument for all $J\ge3$.
When counting candidate supports or vectors, we may include extra choices to obtain an upper bound.

\subsection{Column multiplicities and fixed-weight logical representatives}
For an $m\times n$ binary matrix $H$, write $H_{\cdot j}$ for its $j$th column and define
\[
 \begin{aligned}
 c_H(a)&:=\#\{j\in\{1,\ldots,n\}:H_{\cdot j}=a\},\qquad a\in\mathbb F_2^m,\\
 M(H)&:=\#\{a\in\mathbb F_2^m:c_H(a)=2\}.
 \end{aligned}
\]
Here $c_H(a)$ is the number of occurrences of $a$, and $M(H)$ counts distinct column vectors occurring exactly twice.
The corresponding pairs use disjoint column positions.

For $H_Z$, list its distinct column vectors as $t_1,\ldots,t_s$.
Using the column multiplicity $c_H(a)$ just defined, set
\[
 \begin{aligned}
 s&:=\#\{a\in\mathbb F_2^m:c_{H_Z}(a)>0\},\qquad
 c_i:=c_{H_Z}(t_i)\quad(1\le i\le s),\\
 n_1&:=\#\{i\in\{1,\ldots,s\}:c_i=1\},\qquad
 \Xi:=\sum_{\substack{1\le i\le s\\c_i\ge3}}(c_i-2).
 \end{aligned}
\]
Thus $n_1$ counts columns occurring only once, and $\Xi$ counts all occurrences beyond the first two of each column vector.
At most $3\Xi$ columns have multiplicity at least three.
The number of columns outside the $q=M(H_Z)$ pairs therefore satisfies $n-2q\le n_1+3\Xi$.
Thus, showing that $n_1/n$ and $\Xi/n$ converge to zero in probability will give $q/n\to1/2$.
The same argument applies to $H_X$.
The deterministic count below turns sufficiently many such pairs into a lower bound on each even-weight coefficient after excluding stabilizers.

\begin{lemma}[Logical representatives obtained from equal-column pairs]
\label{lem:pair-count-distance}
Consider a CSS check-matrix pair, and suppose that $H_Z$ contains $q$ equal-column pairs using disjoint column positions.
Let $u_i,v_i$ be the positions of the $i$th pair, and set $b_i=e_{u_i}+e_{v_i}$, where $e_u$ is the standard basis vector with a one in position $u$.
For an integer $0\le r\le q$, the set
\begin{equation}
 \mathcal V_r=
 \left\{\sum_{i\in A}b_i:A\subseteq\{1,\ldots,q\},\ |A|=r\right\}
 \label{eq:fixed-weight-candidates}
\end{equation}
consists of $\binom qr$ kernel vectors of weight $2r$.
If $q-1>\operatorname{rank}H_X$, then, for every integer $1\le r<q$,
\begin{equation}
 \binom{q-2}{r-1}\le
 |\mathcal V_r\setminus S_X|\le\binom qr.
 \label{eq:pair-logical-count}
\end{equation}
The same statements hold with $X$ and $Z$ interchanged.
\end{lemma}
\begin{proof}
Equal columns cancel, so $H_Zb_i=0$.
The vectors $b_1,\ldots,b_q$ have disjoint supports and are therefore linearly independent.
Their sums over $r$-element subsets are therefore distinct and have weight $2r$.
This gives the total number of candidates and the upper bound.

To exclude stabilizers, we compare two candidates of the same weight by replacing $b_i$ with $b_j$.
Their sum is $b_i+b_j$, so if $b_i+b_j\notin S_X$, they cannot both be stabilizers.
We therefore first find such a pair $i,j$.
The vectors $b_1+b_2,\ldots,b_1+b_q$ are linearly independent and span a space of dimension $q-1$.
Since this dimension exceeds $\dim S_X$, there are distinct $i,j$ such that $b_i+b_j\notin S_X$.
Fix such $i,j$.

For any $(r-1)$-element set $T$ avoiding $i,j$, the two vectors
\[
 b_i+\sum_{\ell\in T}b_\ell,\qquad
 b_j+\sum_{\ell\in T}b_\ell
\]
belong to $\mathcal V_r$.
At least one is thus a nontrivial logical representative.
Different choices of $T$ give disjoint pairs of candidates.
There are $\binom{q-2}{r-1}$ choices of $T$, which proves the lower bound.
\end{proof}

To prove concentration, Lemma~\ref{lem:paired-ensemble-lower} lower-bounds the total number of matrix pairs.
Lemmas~\ref{lem:short-type-count} and~\ref{lem:matrix-pair-counting} then bound the choices of $H_Z$ and its commuting partners with prescribed column multiplicities.
Lemma~\ref{lem:gpm-exception-count} combines these counts with block-row parity to bound both singleton and higher-multiplicity columns for every $J\ge3$.
Lemma~\ref{thm:regular-pairing} divides the exceptional-set counts by the total-count lower bound to obtain concentration.
\input{paired_lower_bound.tex}
\input{general_counting_proof.tex}
\input{regular_results.tex}

\subsection{Proof of the coefficientwise lower bound}
\label{app:coefficient-bound-proof}
\begin{proof}[\upshape Proof of Lemma~\ref{lem:enumerator-bound}]
Fix $J/L<\lambda<1/2$ and set $w_{\max,n}=2\lfloor\lambda n\rfloor-2$ and $q=(w_{\max,n}+2)/2=\lfloor\lambda n\rfloor$.
Lemma~\ref{thm:regular-pairing} implies that the event
\[
 M(H_X)\ge q,\qquad M(H_Z)\ge q
\]
has probability tending to one.
Because $\lambda>J/L$, we also have $q-1=w_{\max,n}/2>m$ for all sufficiently large $n$.
On this event, choose any $q$ disjoint equal-column pairs in $H_Z$.
Lemma~\ref{lem:pair-count-distance}, using $\operatorname{rank}H_X\le m$, gives
\[
 A_{X,n}(2r)\ge\binom{w_{\max,n}/2-1}{r-1}
 \qquad(1\le r\le w_{\max,n}/2).
\]
The same argument with $X$ and $Z$ interchanged gives the bound for $A_{Z,n}$ on the same event.
These are all the nonzero coefficients of $z^2(1+z^2)^{w_{\max,n}/2-1}$, so both coefficientwise bounds hold simultaneously.
\end{proof}

%% file: paired_lower_bound.tex
\subsection{Counting a subset of commuting permutation lifts}
\label{app:paired-lower}
\begin{lemma}[A lower bound on the number of matrix pairs in the ensemble]
\label{lem:paired-ensemble-lower}
Fix $J\ge3$ and even $L\ge2J$, and put $n=LP$.
For every positive integer $P$,
\begin{equation}
 |\mathcal G_P(J,L)|\ge(P!)^{JL+L/2}
 =\exp\{(J+\tfrac12)n\log n+O_{J,L}(n)\}.
 \label{eq:paired-ensemble-lower}
\end{equation}
\end{lemma}
\begin{proof}
Fix the block-column pairs $(1,2),(3,4),\ldots,(L-1,L)$.
For each pair $a$, choose arbitrary $Q_a,A_{ia},B_{ja}\in\mathcal P_P$, with $1\le i,j\le J$.
Set
\[
 \begin{aligned}
 ((H_X)_{i,2a-1},(H_X)_{i,2a})&=(A_{ia},A_{ia}Q_a),\\
 ((H_Z)_{j,2a-1},(H_Z)_{j,2a})&=(B_{ja},B_{ja}Q_a).
 \end{aligned}
\]
Every block remains a permutation matrix.
The contribution of block-column pair $a$ to block $(i,j)$ of the check product is
\[
 A_{ia}B_{ja}^{\mathsf T}
 +A_{ia}Q_a(B_{ja}Q_a)^{\mathsf T}=0,
\]
because $Q_aQ_a^{\mathsf T}=I$ and the field is binary.
Thus every constructed pair belongs to $\mathcal G_P(J,L)$.

The odd block columns recover all $A_{ia},B_{ja}$, and
$Q_a=A_{1a}^{-1}(H_X)_{1,2a}$ recovers each $Q_a$.
The construction is therefore injective, including when additional columns coincide across different block-column pairs.
It gives $(P!)^{(2J+1)L/2}$ distinct samples, proving the lower bound.
Its asymptotic form follows from Stirling's formula~\cite[Prop.~B.1]{FlajoletSedgewick2009} and $n=LP$.
This construction counts a subset only; its pairing pattern is not imposed on the ensemble.
\end{proof}

%% file: general_counting_proof.tex
\subsection{Counting dependencies among small sets of columns}
\label{app:general-counting-proof}
To bound exceptional matrix pairs in the ensemble, we first choose $H_Z$ and then a compatible $H_X$.
We use the block-row structure to control column sums, and may overcount the choices of supports or vectors when obtaining upper bounds.
A column occurring only once cannot be used in an equal-column pair.
Any row of $H_X$ containing that position must instead use a larger zero-sum set of columns of $H_Z$.
We decompose such sets into minimal dependencies before counting them.

Identify each column vector with the set of row indices in its support; this set has size $J$.
At most $L$ distinct column vectors contain any given row index.
Write $h_i$ for column $i$.
A nonempty set $S$ of column positions is a \emph{minimal dependent set}, or \emph{circuit}, if $\sum_{i\in S}h_i=0$ and no nonempty proper subset has zero sum.
The support of every nonzero kernel vector is a disjoint union of circuits~\cite[Prop.~9.2.2 and Thm.~9.1.2(vi)]{Oxley2011}.
A circuit of size at least three contains no repeated column vector, since two equal columns would already form a proper zero-sum subset.
Such a circuit can therefore also be viewed as a set of distinct column vectors.

The permutation block structure gives a parity property independent of whether $J$ is even or odd.
Each column has one nonzero entry in each of the $J$ block rows, each containing $P$ ordinary rows.
The sum of $r$ columns therefore has weight congruent to $r$ modulo two within every block row.
Consequently, every circuit has even size.
Moreover, a nonzero sum has weight at least $J$ if $r$ is odd, and at least two if $r$ is even.
In particular, every nonzero sum has weight at least two because $J\ge3$.
We use both consequences below.

To compare with the total-count lower bound of Lemma~\ref{lem:paired-ensemble-lower}, we track powers of $n$ and absorb factors of the form $\exp\{O_{J,L}(n)\}$.
A decrease of $\varepsilon n$ in the exponent of $n$ reduces the count by a factor $\exp(-\varepsilon n\log n)$.

\subsubsection{Choosing column vectors contained in short dependencies}
If many small zero-sum sets occur in addition to equal-column pairs, their column vectors cannot be chosen freely.
Let $A$ be the set of distinct column vectors belonging to at least one circuit of size between four and $2J$, and put $a=|A|$.
Circuits other than equal-column pairs have size at least four; any such circuit containing a vector outside $A$ has size at least $2J+2$.

\Needspace{6\baselineskip}
\begin{lemma}[An upper bound on the number of sets $A$]
\label{lem:short-type-count}
For the set $A$ defined above, fix $a=|A|$ and require each vector in $A$ to have one nonzero entry in each block row.
If at most $L$ column vectors in $A$ contain any given ordinary row index, the number of possible sets $A$ is at most
\begin{equation}
 \exp\{O_{J,L}(n)\}\,n^{Ja/2}.
 \label{eq:short-type-count}
\end{equation}
\end{lemma}
\begin{proof}
Build $A$ by adding one circuit at a time.
We show that a step introducing $j$ new column vectors has at most $O_{J,L}(n^{Jj/2})$ choices.
Multiplying over the steps will give the result because the total number of new column vectors is $a$.

To reach every vector in $A$, use a graph to organize the additions.
Its vertices are the circuits of sizes between four and $2J$, and two are adjacent when they share a column vector.
Within each connected component, start with one circuit and successively add circuits that share a previously selected vector and contain a new vector.
Suppose the circuit being added has size $\ell\le 2J$, with $j\ge1$ new column vectors and $k=\ell-j$ previously selected vectors.
Here $k$ counts previously selected vectors; it is a local variable, distinct from the number of logical qubits denoted by $k$ earlier.

For one addition, we first choose which previously selected vectors to reuse and then choose the row indices supporting the new vectors.
To count the first choice, use a second graph whose vertices are the previously selected column vectors, joining two when their supports intersect.
Suppose the $k$ previously selected vectors used at this step have $c$ connected components in their induced intersection graph.
The intersection graph on all previously selected vectors has maximum degree at most $J(L-1)$.
For each component, there are $O(n)$ choices for its first vertex and only a constant number of choices for the remaining vertices, because its size is bounded.
After specifying the component sizes, there are therefore $O_{J,L}(n^c)$ choices.

Fix these previously selected vectors and consider the sum that the new vectors must cancel.
Let $b_i$ be the weight of the sum of the column vectors in component $i$.
Different components have disjoint row supports, so the weight of the sum of all $k$ vectors is $b=\sum_i b_i$.
Each component is a proper subset of the circuit being added; hence $b_i>0$.
Otherwise that component would itself sum to zero, contradicting minimality.

The $j$ new columns contain $Jj$ ones in total.
At least $b$ of them must cancel the nonzero entries in the sum of the previously selected vectors.
Their row indices are already determined.
In any row unused by the $k$ previously selected vectors, the new vectors must contribute an even number of ones.
Each newly used row therefore contains at least two of the remaining $Jj-b$ ones.
At most $(Jj-b)/2$ new row indices can be chosen freely.
Rows already in the supports of the selected $k$ vectors are known, and the assignments of ones among these rows and among the new columns have only a constant number of possibilities.
Multiplying the choices of previously selected vectors and new row indices gives at most
\begin{equation}
 O_{J,L}\!\left(n^{\,c+(Jj-b)/2}\right)
 \label{eq:regular-growth-exponent}
\end{equation}
choices for this addition.
The initial step $k=0$ satisfies the same bound with $c=b=0$.

Each component is a proper subset of the circuit, so its column sum is nonzero.
The block-row parity property gives $b_i\ge2$ for every component and hence $b\ge2c$.
The exponent in~\eqref{eq:regular-growth-exponent} is therefore
$Jj/2+(c-b/2)\le Jj/2$.
This argument applies to every $J\ge3$.

Every step has at most $O_{J,L}(n^{Jj/2})$ choices.
There are at most $a$ steps.
Specifying the circuit size, the number of new vectors, and whether a new component starts at each step gives at most $e^{O_{J,L}(a)}$ descriptions in total.
Processing each connected component of the first graph in this way obtains all vectors in $A$.
The choices of previously selected vectors are already counted in~\eqref{eq:regular-growth-exponent}, so no additional factorial for their order is needed.
Retain only partial sets whose vectors have one nonzero entry in each block row and with at most $L$ distinct vectors containing any ordinary row index.
These restrictions preserve the parity property and degree bound throughout, and can only reduce the number of choices.
Since $\sum j=a$, we obtain~\eqref{eq:short-type-count}.
\end{proof}

\subsection{Counting \texorpdfstring{$H_Z$ and compatible matrices $H_X$}{HZ and compatible matrices HX}}
The bound on the number of sets $A$ lets us count the choices for $H_Z$, followed by the compatible matrices $H_X$.
The number $s$ of distinct column vectors and the multiplicity parameters satisfy $\Xi=n+n_1-2s$.
Indeed, summing $c_i-2$ over all types gives a contribution of $-1$ from each type occurring once, zero from each type occurring twice, and $c_i-2$ from every other type.
Partition the columns occurring once according to whether their column vector belongs to $A$:
\[
 \begin{aligned}
 u&:=\#\{i\in\{1,\ldots,s\}:c_i=1,\ t_i\in A\},\\
 v&:=\#\{i\in\{1,\ldots,s\}:c_i=1,\ t_i\notin A\}.
 \end{aligned}
\]
Then $n_1=u+v$ and $u\le a$.
We distinguish these cases because they impose different minimum circuit sizes.

\begin{lemma}[Upper bounds for counting a matrix pair in two stages]
\label{lem:matrix-pair-counting}
For specified $s,a,n_1,\Xi$, the number of block arrays $H_Z$ is at most
\begin{equation}
 \exp\{O_{J,L}(n)\}\,n^{n+(J-1)s-(J-2)a/2}.
 \label{eq:regular-hz-count}
\end{equation}
For fixed $H_Z$, decompose the support of each $X$ row into disjoint circuits, choosing one decomposition by a fixed rule, such as lexicographic order.
The number of compatible block arrays $H_X$ whose total number of circuit uses is $C_X$ is at most $\exp\{O_{J,L}(n)\}n^{C_X}$.
\end{lemma}
\begin{proof}
First choose $A$ using Lemma~\ref{lem:short-type-count}.
There are at most $\binom{\binom mJ}{s-a}\le\exp\{O_{J,L}(n)\}n^{(J-1)(s-a)}$ choices for the remaining $s-a$ distinct vectors, at most $L^s$ assignments of their multiplicities, and at most $n!$ assignments to column positions.
The remaining vectors are chosen as an unordered set, so their count includes division by $(s-a)!$.
This changes the exponent of $n$ per type from $J$ to $J-1$.
The bound holds even when $t=s-a$ is not proportional to $n$.
To see this, take an integer constant $C=C_{J,L}\ge1$ such that $\binom mJ\le Cn^J$.
For $1\le t\le n$,
$\binom{Cn^J}{t}\le n^{(J-1)t}(eCn/t)^t$.
Since $t\log(n/t)\le n/e$, the last factor is $\exp\{O_{J,L}(n)\}$.
For $t=0$, there is one choice.
Multiplying by $n^{Ja/2}$ for $A$ and by $n!$ for the positions gives the exponent $n+(J-1)s-(J-2)a/2$ in~\eqref{eq:regular-hz-count}.

An $X$ row commutes with $H_Z$ precisely when the columns indexed by its support sum to zero.
Decompose this support into disjoint circuits~\cite[Prop.~9.2.2 and Thm.~9.1.2(vi)]{Oxley2011}, using the fixed rule in the statement.
For any fixed size $\ell\le L$, there are at most $O_{J,L}(n)$ circuits.
Indeed, the graph joining two column positions when their supports intersect has maximum degree at most $J(L-1)$.
If a circuit were disconnected in this graph, each component would have zero column sum, contradicting minimality.
It is therefore connected.
After choosing its first position in $n$ ways, there are only a constant number of choices for the remaining at most $L-1$ positions.

Let $C_X$ be the total number of circuit uses in all $X$ rows.
Once the circuit sizes used in each row are specified, the number of candidate matrices $H_X$ is at most $\exp\{O_{J,L}(n)\}n^{C_X}$.
Here $C_X$ counts uses, rather than distinct circuits: a circuit used in two different rows is counted twice.
Each use has $O_{J,L}(n)$ choices, which explains the exponent $C_X$.
There are only a constant number of size patterns per row, and summing over their choices in all rows contributes another factor $e^{O_{J,L}(n)}$.
\end{proof}

\subsection{An upper bound on exceptional matrix pairs}
\label{app:gpm-exception-proof}
We now bound the circuit-use count $C_X$ in Lemma~\ref{lem:matrix-pair-counting}.
The block-row parity property ensures that a circuit other than an equal-column pair uses at least four columns.
A column occurring once must lie in such a circuit, which decreases $C_X$ and hence the number of compatible matrices.

\begin{lemma}[Counting matrix pairs with exceptional column multiplicities]
\label{lem:gpm-exception-count}
Let $J\ge3$.
For specified $s,a,n_1,\Xi$, the number of matrix pairs in the ensemble is at most
\begin{equation}
 \exp\{O_{J,L}(n)\}\,
 n^{(J+1/2)n-(J-1)\Xi/2-n_1/(2J+2)}.
 \label{eq:gpm-pair-upper}
\end{equation}
\end{lemma}
\begin{proof}
Decompose each $X$-row support into circuits by the fixed rule in Lemma~\ref{lem:matrix-pair-counting}.
For one circuit of size $\ell$, let $u_\ell,v_\ell$ count its columns corresponding to the singleton classes $u,v$.
We claim that
\[
 1\le\frac\ell2-\frac{u_\ell}{4}
       -\left(\frac12-\frac1{2J+2}\right)v_\ell.
\]
For $\ell=2$, both columns are equal, so $u_\ell=v_\ell=0$ and equality holds.
If $v_\ell=0$ and $\ell\ge4$, the right-hand side is at least $\ell/4\ge1$.
If $v_\ell>0$, the circuit contains a vector outside $A$, so its even size is at least $2J+2$.
Since the coefficient of $v_\ell$ in the subtraction is at least $1/4$ and $u_\ell+v_\ell\le\ell$, the right-hand side is at least $\ell/(2J+2)\ge1$.

Summing over all circuit uses gives
\begin{equation}
 C_X\le Jn/2-Ju/4-J\left(1/2-1/(2J+2)\right)v.
 \label{eq:gpm-completion-budget}
\end{equation}
Here the circuit sizes sum to $Jn$, and each singleton column position occurs in exactly $J$ row supports of $H_X$.
Thus the sums of $u_\ell,v_\ell$ are $Ju,Jv$.

Multiply the two-stage counts in Lemma~\ref{lem:matrix-pair-counting}, using $s=(n+u+v-\Xi)/2$.
The exponent of $n$ is at most
\begin{align*}
 &(J+\tfrac12)n-\tfrac{J-1}{2}\Xi
       -\tfrac{J-2}{2}a+\tfrac{J-2}{4}u-\tfrac{v}{2J+2}\\
 &\quad\le(J+\tfrac12)n-\tfrac{J-1}{2}\Xi
       -\tfrac{J-2}{4}a-\tfrac{v}{2J+2}\\
 &\quad\le(J+\tfrac12)n-\tfrac{J-1}{2}\Xi-\tfrac{n_1}{2J+2}.
\end{align*}
The first inequality uses $u\le a$.
For the second, use $a\ge u$, $n_1=u+v$, and $(J-2)/4\ge1/(2J+2)$ for $J\ge3$.
There are only polynomially many choices for $u,v$ and the circuit-use counts, so summing over them is absorbed into $\exp\{O_{J,L}(n)\}$.
This proves~\eqref{eq:gpm-pair-upper} for all $J\ge3$.
\end{proof}

%% file: regular_results.tex
\subsection{From exceptional counts to concentration of columns}
\label{sec:regular-counting-results}
The preceding bounds control the fraction of columns occurring once or at least three times.
Together they imply $M(H)/n\to1/2$ in probability.

\begin{lemma}[Concentration on equal-column pairs]
\label{thm:regular-pairing}
Fix an integer $J\ge3$ and an even integer $L\ge2J$, and put $n=LP$ as $P\to\infty$.
For a pair sampled uniformly from the ensemble $\mathcal G_P(J,L)$, the number $M(H)$ of distinct column vectors occurring exactly twice satisfies
\begin{equation}
 \frac{M(H_X)}n\xrightarrow{\Pr}\frac12,\qquad
 \frac{M(H_Z)}n\xrightarrow{\Pr}\frac12.
 \label{eq:regular-pairing}
\end{equation}
\end{lemma}
\begin{proof}
Let $n_1(H)$ and $\Xi(H)$ denote the two multiplicity parameters defined at the beginning of the section, evaluated for $H$.
Under the uniform distribution, an event's probability is the number of matrix pairs satisfying it divided by the total number of pairs.
Use the permutation-lift lower bound of Lemma~\ref{lem:paired-ensemble-lower} for the denominator.
For the numerator, the unified exceptional-count bound in Lemma~\ref{lem:gpm-exception-count} decreases the exponent of $n$ by $n_1/(2J+2)$ or $(J-1)\Xi/2$.
The remaining parameters have only polynomially many possible values, so summing over them contributes a factor absorbed into $\exp\{O(n)\}$.
The common leading term $(J+1/2)n\log n$ cancels in the logarithm of the ratio.
Thus, for any fixed $\varepsilon,\eta>0$,
\begin{align*}
 \Pr\{n_1(H)\ge\varepsilon n\}
 &\le\exp\{-\tfrac{\varepsilon}{2J+2}n\log n+O_{J,L,\varepsilon}(n)\}\longrightarrow0,\\
 \Pr\{\Xi(H)\ge\eta n\}
 &\le\exp\{-\tfrac{J-1}{2}\eta n\log n+O_{J,L,\eta}(n)\}\longrightarrow0.
\end{align*}
Here $H$ can be either $H_X$ or $H_Z$.
Both $n_1(H)/n$ and $\Xi(H)/n$ converge to zero in probability.

Partition the columns by multiplicity to obtain
\[
 n=n_1(H)+2M(H)+\sum_{c_i\ge3}c_i.
\]
For $c_i\ge3$, the inequality $c_i\le3(c_i-2)$ bounds the last sum by $3\Xi(H)$.
Hence the number of columns not occurring exactly twice satisfies
\[
 0\le n-2M(H)\le n_1(H)+3\Xi(H)=o_{\Pr}(n),
\]
which proves $M(H)/n\to1/2$.
The same estimates hold for both matrices.
A union bound on their failure probabilities gives the two convergences simultaneously.
\end{proof}

%% file: weight_two_appendix.tex
\section{An alternative proof using weight-two logical operators}
\label{app:weight-two-proof}
Weight-two logical operators give an alternative proof of the main theorem's conclusion.
We first give a condition for distance two in terms of the number of equal-column pairs and the check-matrix rank, then apply it to uniform sampling using the concentration result in Lemma~\ref{thm:regular-pairing}.

\begin{lemma}[A rank condition for distance two]
\label{lem:weight-two-rank}
Suppose every column of a CSS check-matrix pair $(H_X,H_Z)$ is nonzero.
If $H_Z$ contains $q>\operatorname{rank}H_X$ equal-column pairs using disjoint column positions, then $d_X=2$.
The same statement holds with $X$ and $Z$ interchanged.
\end{lemma}
\begin{proof}
If positions $i,j$ have equal columns, then $H_Z(e_i+e_j)=0$.
The weight-two vectors from the $q$ disjoint pairs are linearly independent, so they cannot all belong to $\operatorname{row}H_X$.
At least one is a nontrivial $X$-type logical representative, giving $d_X\le2$.
Since every column is nonzero, no weight-one vector $e_i$ belongs to $\ker H_Z$.
Thus $d_X=2$.
Interchanging the matrices gives the other statement.
\end{proof}

\begin{lemma}[The distance of a uniformly sampled code]
\label{lem:weight-two-uniform}
Fix an integer $J\ge3$ and an even integer $L>2J$, and put $n=LP$ as $P\to\infty$.
For a uniformly sampled pair in $\mathcal G_P(J,L)$, with high probability,
\begin{equation}
 d_X=d_Z=2.
 \label{eq:weight-two-uniform}
\end{equation}
\end{lemma}
\begin{proof}
Lemma~\ref{thm:regular-pairing} gives $M(H)/n\to1/2$ in probability for both matrices.
Since $m/n=J/L<1/2$, the event $M(H)>m$ on both sides has probability tending to one.
On this event, the columns occurring exactly twice in $H_Z$ give disjoint pairs, and their number is greater than $m\ge\operatorname{rank}H_X$.
All columns are nonzero because their weight is $J\ge3$.
Lemma~\ref{lem:weight-two-rank} therefore gives $d_X=2$.
Interchanging the matrices also gives $d_Z=2$.
\end{proof}

For every fixed $\delta>0$, we have $2<\delta n$ once $n>2/\delta$.
Lemma~\ref{lem:weight-two-uniform} gives $d_n=2<\delta n$ with high probability, proving the conclusion of the main theorem independently of its main-text proof.